\documentclass[a4paper,12pt]{article}
\pdfoutput=1
\usepackage[margin=1in]{geometry}
\usepackage{graphicx,xcolor}
\usepackage{mathtools,braket,blkarray}
\allowdisplaybreaks
\usepackage{amssymb,amsfonts,bbm,relsize}
\usepackage[width=0.9\textwidth,footnotesize]{caption}
\usepackage[font=scriptsize]{subcaption}
\usepackage{cite,hyperref,url}
\hypersetup{pageanchor=false}
\usepackage{booktabs,multirow,makecell}
\usepackage[utf8]{inputenc}
\usepackage{cleveref}
\usepackage{amsmath}
\usepackage{breqn}

\newcommand{\vev}[1]{\left< #1 \right>}
\newcommand{\abs}[1]{| #1 |}

\newcommand{\ka}[1]{K\"{a}hler}

\begin{document}

\begin{titlepage}
\begin{center}
{\bf\Large  
Starobinsky-like inflation and soft-SUSY breaking 
  } \\[12mm]
Stephen~F.~King$^{\star}$%
\footnote{E-mail: \texttt{king@soton.ac.uk}},
Elena~Perdomo$^{\star}$%
\footnote{E-mail: \texttt{e.perdomo-mendez@soton.ac.uk}}
\\[-2mm]

\end{center}
\vspace*{0.50cm}
\centerline{$^{\star}$ \it
School of Physics and Astronomy, University of Southampton,}
\centerline{\it
SO17 1BJ Southampton, United Kingdom }

\vspace*{1.20cm}

\begin{abstract}
{\noindent
We study a version of Starobinsky-like inflation in no-scale supergravity (SUGRA) where a Polonyi term in the hidden sector
breaks supersymmetry (SUSY) after inflation, providing a link between 
the gravitino mass and inflation. We extend the theory to the visible sector and calculate the 
soft-SUSY breaking parameters depending on the modular weights in the superpotential and choice of \ka{1} potential.
We are led to either no-scale SUGRA or pure gravity mediated SUSY breaking patterns, but with 
inflationary constraints on the Polonyi term setting a strict upper bound on the gravitino mass $m_{3/2}<10^3$ TeV.
Since gaugino masses are significantly lighter than $m_{3/2}$,
this suggests that SUSY may be discovered at the LHC or FCC.
}
\end{abstract}
\end{titlepage}
\section{Introduction}

Inflation \cite{Guth:1980zm,Linde:1981mu,Mukhanov:1981xt,Albrecht:1982wi,Linde:1983gd,Linde:2007fr} is well known to 
solve the flatness and horizon problems, diluting cosmological relics and providing an origin of cosmological fluctuations. 
In slow-roll inflation \cite{Linde1990,LythRiotto1999}, the inflaton rolls along a quite flat potential and 
inflation end as it falls into some basin. Inflation is supported and constrained by current observational data \cite{Ade:2015lrj},
which measures a spectral index $n_s\approx 0.96 \pm  0.007$ and low tensor-to-scalar ratio $r< 0.08$,
excluding the simplest chaotic models
based on polynomial potentials such as $\phi^2$ or $\phi^4$  \cite{MartinRingevalVennin2014}.
Surviving models include Starobinsky inflation
\cite{Mukhanov:1981xt,R2}, Higgs inflation \cite{HI} and related models \cite{others}, and low scale hybrid inflation \cite{Copeland:1994vg}.

Supersymmetry (SUSY) may be naturally combined with inflation since it allows 
better control over the high energy dynamics of scalars
 \cite{Ellis:1982dg,Ellis:1982ed,Ellis:1982ws}.
 SUSY inflation is also motivated by the Lyth bound \cite{Lyth:1984yz} on the low tensor-to-scalar ratio,
 which prefers a scale of inflation below the Planck scale.
Since inflation is sensitive to UV scales,
it is necessary to consider supergravity (SUGRA) inflation, as in e.g.
\cite{YanagidaSUGRA,AntushcSUGRAinflation,KalloshSUGRAinflation}. In no-scale SUGRA \cite{Ellis:1984bf}, the K\"ahler potential takes a logarithmic form which circumvents the $\eta$ problem. 
Alternatives to no-scale SUGRA have also been proposed which also address the $\eta$ problem
based on a Heisenberg symmetry
\cite{BG,HG} or a shift symmetry \cite{Y,Davis:2008fv,klor}
(see also \cite{NoScaleInflation}).

It has been shown by Ellis, Nanopoulos, Olive (ENO) 
that no-scale SUGRA can behave like a Starobinsky inflationary model \cite{Ellis:2013xoa,Ellis:2013nxa,Ellis:2013nka}.
However, in this approach, a term with constant modular weight 
is used to break SUSY, and there is no connection between inflation and SUSY breaking.
Recently we considered the above ENO model, but 
with a linear Polonyi term added to the superpotential \cite{Romao:2017uwa}. 
The purpose of adding this term was to provide
an explicit mechanism for breaking SUSY in order to provide a link between inflation and SUSY breaking. 
Indeed we showed that inflation requires a strict upper bound for the gravitino mass $m_{3/2}< 10^3$ TeV \cite{Romao:2017uwa}.

In the present paper we show how the Polonyi-extended ENO model 
may be generalised to include the fields in the 
visible sector of the minimal supersymmetric standard model (MSSM).
Such a generalisation has been done for the ENO model \cite{Ellis:2013xoa,Ellis:2013nxa,Ellis:2013nka}
and we perform a similar analysis for the Polonyi-extended ENO model.
We calculate the 
soft-SUSY breaking parameters depending on the modular weights in the superpotential and choice of \ka{1} potential 
and we are led to new phenomenological possibilities for supersymmetry (SUSY) breaking, based on generalisations of no-scale SUSY breaking and pure gravity mediated SUSY breaking.
The Polonyi-extended ENO model is especially interesting to consider 
because of the upper bound on the gravitino mass discussed in the previous paragraph which allows the much lighter 
gauginos to be discovered in future collider experiments. This motivates the present investigation of the soft 
SUSY breaking parameters, which could form the basis for future phenomenological studies.

The layout of the remainder of the paper is as follows.
In section~\ref{sec:Kahler_function} we discuss the hidden sector of the supergravity theory where inflation takes place.
In section~\ref{visible} we discuss the visible sector of the supergravity theory and show how the MSSM matter and Higgs fields may be included. In section~\ref{sec:soft} we discuss the supergravity scalar potential, showing how inflation emerges from the hidden sector and soft-SUSY breaking parameters emerge from the full theory including the visible sector, leading to new examples of  
no-scale SUSY breaking and pure gravity mediated SUSY breaking.
Section~\ref{conclusion} concludes the paper.

\section{The Hidden Sector}
\label{sec:Kahler_function}

In general supergravity theory,
the tree-level supergravity scalar potential can be found using the \ka{1} function $G$, which is given
in terms of the \ka{1} potential $K$ and the superpotential $W$ as,
\begin{equation}
G=\frac{K}{M_P^2} + \ln \abs{\frac{W}{M_P^3}} +\ln \abs{\frac{W}{M_P^3}}^*. 
		\label{eq:ka_function}
\end{equation}
The effective scalar potential is then given by,
\begin{equation}
	V=e^G \left[\frac{\partial G}{\partial \phi_i} K_{ij^*} \frac{\partial G}{\partial \phi_{j^*}}-3\right] M_P^4,
	\label{eq:scalar_potential}
\end{equation}
where $K_{ij^*}$ is the inverse of the \ka{1} metric $K^{ij^*}\equiv \partial^2 K /\partial \phi_i \partial \phi^*_{j}$. When, at the minimum of the scalar potential, some of the hidden sector fields acquire VEVs in such a way that at least one of their auxiliary fields, $F^i$, is non-vanishing, then SUSY is spontaneously broken and soft SUSY-breaking terms are generated in the observable sector. The gravitino becomes massive and its mass 
\begin{equation}
	m_{3/2}^2=e^{G}=e^{K/M_P^2}\frac{\abs{W}^2}{M_P^4}
	\label{eq:gravitino_mass}
\end{equation} 
sets the overall scale of the soft parameters. 
In general, the \ka{1} potential $K$ and the superpotential $W$ involve superfields in the hidden and visible sectors.
The hidden sector superfields are gauge singlets and do not have Yukawa couplings to the charged fields in the visible sector,
coupling only indirectly to them via Planck suppressed operators.

The simplest no-scale \ka{1} potential in the hidden sector
is given by two complex fields $(T, \phi)$, where T is a modulus field while $\phi$ is the field responsible for inflation and SUSY breaking. The \ka{1} potential in the hidden sector takes the form
\begin{equation}
	K_{hid}(\phi, T)= -3 M_P^2 \ln \left(\frac{T+T^*}{M_P}-\frac{\abs{\phi}^2}{3M_P^2}\right),
	\label{eq:ka_potential}
\end{equation}
where $M_P$ is the reduced Planck scale. 

It was found in \cite{Ellis:2013xoa} this \ka{1} potential together with the Wess-Zumino superpotential~\cite{Wess:1974tw, Croon:2013ana} can lead to the Starobinsky-like inflationary potential. When the modulus field $T$ is fixed with a vacuum expectation value of $\vev{Re\ T}=1/2$ and $\vev{Im\ T}=0$, the no-scale \ka{1} potential together with the Wess-Zumino superpotential is equivalent of an $R+R^2$ model of gravity, in which Starobinsky inflation emerges at a particular point in parameter space \cite{Ellis:2013nxa}. A simple modification to this superpotential has been done in \cite{Romao:2017uwa}, adding the Polonyi term to provide an explicit and simple mechanism for supersymmetry breaking at the end of inflation. The Wess-Zumino superpotential~\cite{Wess:1974tw}
in the hidden sector, with quadratic and trilinear terms, together with a linear Polonyi term looks like
\begin{equation}
	W_{hid}(\phi)=m^2\phi +\frac{\mu}{2} \phi^2 -\frac{\lambda}{3}\phi^3
	\label{eq:Wess-Zumino-Polonyi_sup}.
\end{equation}

In the following, it is convenient to introduce the change of variables~\cite{Ellis:1984bm}
\begin{equation}
	T=\frac{ M_P}{2}\left(\frac{1-\frac{y_2}{\sqrt{3} M_P}}{1+\frac{y_2}{\sqrt{3} M_P}}\right), \quad \phi=\left(\frac{y_1}{1+\frac{y_2}{\sqrt{3} M_P}}\right), 
	\label{eq:T_variables}
\end{equation}
with the inverse relations
\begin{equation}
	y_1=\left(\frac{2\phi}{1+2T/M_P}\right), \quad y_2=\sqrt{3} M_P\left(\frac{1-2T/ M_P}{1+2T/ M_P}\right).
	\label{eq:y_variables}
\end{equation}
After this change of variables, the hidden sector superpotential becomes 
\begin{equation}
 W_{hid}(y_1,y_2)= \left(1+\frac{y_2}{\sqrt{3} M_P}\right)^{-3} \widetilde{W}_{hid}(y_1,y_2), 
	\label{Whid}
\end{equation}
where the rescaled superpotential is
\begin{equation}
	\widetilde{W}_{hid}(y_1,y_2)=m^2 y_1 \left(1+\frac{y_2}{\sqrt{3} M_P}\right)^2 +\frac{\mu}{2} y_1^2 \left(1+\frac{y_2}{\sqrt{3} M_P}\right) -\frac{\lambda}{3} y_1^3.
	\label{eq:W_y1y2}
\end{equation}

The \ka{1} potential in the hidden sector becomes 
\begin{equation}
K_{hid}(y_1,y_2)=\widetilde{K}_{hid}(y_1,y_2)+3\ln \left(\abs{1+\frac{y_2}{\sqrt{3}M_P}}^2\right)
\end{equation}
where 
\begin{equation}
\widetilde{K}_{hid}(y_1,y_2)=
-3  M_P^2 \ln \left(1-\frac{\abs{y_1}^2+\abs{y_2}^2}{3  M_P^2}\right), 
\label{eq:ka_potential_y_variables}
\end{equation}
Note that the combination of $\widetilde{W}_{hid}$ and $\widetilde{K}_{hid}$
is equivalent to using ${W}_{hid}$ and ${K}_{hid}$, since the physical quantities are given by the \ka{1} function $G$~\ref{eq:ka_function} which is the same in both cases 
(the extra term in the \ka{1} potential cancels with an opposite term coming from the superpotential).
Therefore, we use the symmetric representation of the \ka{potential}
$\widetilde{K}_{hid}$ in Eq.~\ref{eq:ka_potential_y_variables} and the rescaled superpotential $\widetilde{W}_{hid}$
in Eq.\ref{eq:W_y1y2} in the following.

This superpotential can reproduce the Starobinsky model for the real part of $y_1$ and fixed $\vev{y_2}=0$, when a suitable stabilizing term $\propto y^4_2$ is added to the no-scale \ka{1} potential \cite{Ellis:2013nxa, Ellis:1984bm, Kallosh:2013xya}: 
\begin{equation}
	\widetilde{K}_{hid}(y_1,y_2)=-3 M_P^2 \ln \left(1-\frac{\abs{y_1}^2+\abs{y_2}^2}{3 M_P^2}+ \frac{\abs{y_2}^4}{\Lambda^2 M_P^2}\right),
	\label{eq:ka_stabilizing_term}
\end{equation}
with $\Lambda \lesssim 0.1 M_P$, as discussed in \cite{Ellis:2013nxa}. In section~\ref{visible}, we ignore the term $\propto y_2^4$ for simplicity, but it should be kept in mind that it is necessary to stabilize the field $\vev{y_2}=0$.

The exact Starobinsky potential is obtained when dropping the Polonyi term, $m=0$, fixing $\vev{y_2}=0$ and using the relationship $\lambda=\mu/\sqrt{3} M_P$ \cite{Ellis:2013nxa}. To quantify how much the Starobinsky limit deviates when including the Polonyi term, we use the parameter $b=m^2/3\lambda M_P^2$ while keeping the relation  $\lambda=\mu/\sqrt{3} M_P$, which gives 
\begin{equation}
\mu^{-1}\widetilde{W}_{hid}(y_1,y_2)=\sqrt{3}b M_P y_1 \left(1+\frac{y_2}{\sqrt{3}M_P}\right)^2 +\frac{y_1^2}{2}  \left(1+\frac{y_2}{\sqrt{3}M_P}\right) -\frac{y_1^3}{3 \sqrt{3}M_P}.
	\label{eq:W_inflation_limit}
\end{equation}
For non-zero $b$, SUSY is broken and the gravitino mass in Eq.\ref{eq:gravitino_mass}
becomes non-zero at the end of inflation, as was shown in \cite{Romao:2017uwa}. As discussed in~\cite{Romao:2017uwa}, there is an upper limit on the parameter $\abs{b}$ in order to have a viable inflationary scenario, suggesting a gravitino mass $m_{3/2}<10^3$ TeV with favoured values of $m_{3/2}\sim \mathcal{O}(1)$ TeV.
A more quantitative discussion (which we do not repeat here) can be found in \cite{Romao:2017uwa}, where it is shown that 
this limit comes from the two crucial dimensionless observables: the tensor-to-scalar ratio, $r$, and the scalar tilt, $n_s$, given by the Planck satellite~\cite{Ade:2015lrj}. Furthermore, the scalar amplitude observable, $A_s$, is sensitive to the overall scale of the potential, $i.e$ to the parameter $\mu$ in Eq.~\ref{eq:W_inflation_limit}. It is shown there  \cite{Romao:2017uwa} that, in the Starobinsky limit when $\lambda=\mu/\sqrt{3}M_P$, the bilinear mass term parameter becomes $\mu \simeq 10^{-5}M_P$ and we use these values in the following computations.

\section{The Visible Sector}
\label{visible}
We now extend the hidden sector inflation model in the previous section
to include the matter fields in the visible sector, such as those of the 
minimal supersymmetric standard model (MSSM), which includes the 
Standard Model (SM) quarks and leptons 
together with their supersymmetric partners, written generically in terms of matter and Higgs superfields  
$y_{vis}=\hat{Q}_i,\hat{U}_i^c,\hat{D}_i^c,\hat{L}_i,\hat{E}_i^c,\hat{H}_u, \hat{H}_d$.
The visible sector superfields carry gauge charges under the SM gauge group, unlike the hidden sector
supefields which are gauge singlets.
We consider two possibilities for extending the hidden sector supergravity theory of the previous section 
to include such visible sector superfields. 

Motivated by the no-scale approach~\cite{No-scale}, the first possibility (Case I) is to
embed the visible sector matter superfields within the logarithm in the \ka{1} potential (Case I), such that 
\begin{equation}
K_I=-3M_P^2 \ln \left(1-\frac{\abs{y_1}^2+\abs{y_2}^2+\abs{y_{vis}}^2}{3M_P^2}\right), \quad \text{Case I}.
\label{eq:Ka_SM_inside}
\end{equation}
Another possibility (Case II) we also explore is to have the visible sector superfields outside the logarithm in the \ka{1} potential via minimal kinetic terms
\begin{equation}
K_{II}=-3M_P^2\ln \left(1-\frac{\abs{y_1}^2+\abs{y_2}^2}{3M_P^2}\right) +\abs{y_{vis}}^2, \quad \text{Case II}.
\label{eq:Ka_SM_outside}
\end{equation}
Furthermore, in the same spirit as in \cite{Ellis:2013nka}, we assume that the superpotential for the
visible sector superfields has the form 
\begin{equation}
	W_{vis}=W_2(y_{vis})\left(1+\frac{y_2}{\sqrt{3}M_P}\right)^\beta+W_3(y_{vis})\left(1+\frac{y_2}{\sqrt{3}M_P}\right)^\alpha,
	\label{eq:SM_superpotential}
\end{equation}
where $\alpha$ and $\beta$ are modular weights and $W_{2,3}(y_{vis})$ are bi/trilinear 
parts of the superpotential of the MSSM in terms of the visible sector superfields $y_{vis}$. The total superpotential is then given by 
\begin{equation}
W=\widetilde{W}_{hid}+W_{vis},
\label{W}
\end{equation}
where $\widetilde{W}_{hid}$ was given in Eq.\ref{eq:W_inflation_limit} and $W_{vis}$ is given in Eq.\ref{eq:SM_superpotential}.

We want to make the connection between the Wess-Zumino-Polonyi superpotential~\ref{eq:W_inflation_limit}, which has a parameter space in which the Starobinsky inflation is recovered, and the soft supersymmetry-breaking parameters after the MSSM superfields are included. The main difference between the model of  \cite{Romao:2017uwa} which we are developing here,
and that proposed in \cite{Ellis:2013nka} is that, in the present case,
supersymmetry is broken through the Polonyi term while in  \cite{Ellis:2013nka} a term with constant modular weight 3 term
was used to break SUSY. In the present model there is a constraint from inflation in how big the parameter $b$ (accounting for the Polonyi term) can be, leading to an upper bound on the gravitino mass, which sets the SUSY breaking scale.
The present model therefore suggests a more constrained region of 
parameter space which may be confronted with LHC, Higgs and dark matter constraints. 

The main difference between the two cases I and II is that in the pure no-scale case \ref{eq:Ka_SM_inside}, the soft supersymmetry-breaking mass squared $m_0$ is zero while in the latter case \ref{eq:Ka_SM_outside}, $m_0$ is different from zero and equal to the gravitino mass, as we will show in section~\ref{sec:soft}.

\section{Potential and soft-SUSY breaking parameters}
\label{sec:soft}
We can compute the supergravity scalar potential from Eqs.~\ref{eq:ka_function} and \ref{eq:scalar_potential},
using either Eq.\ref{eq:Ka_SM_inside} (case I) or Eq.\ref{eq:Ka_SM_outside} (case II),
with the total superpotential in Eq.\ref{W}.
At the minimum of the potential
some of the hidden sector fields, $y_1$ and $y_2$, acquire VEVs such that the F term is non-vanishing, then supersymmetry is spontaneously broken and soft supersymmetry-breaking terms are generated in the observable sector, $y_{vis}$.
This is just the usual gravity mediated SUSY breaking mechanism.
However the special forms of superpotential and \ka{1} potential here lead to a special form of SUSY breaking,
referred to as either no-scale SUSY breaking (case I) or 
pure gravity mediated SUSY breaking (case II).

\subsection{Hidden sector potential and inflation}
\begin{figure}
	\centering
	\includegraphics[scale=0.25]{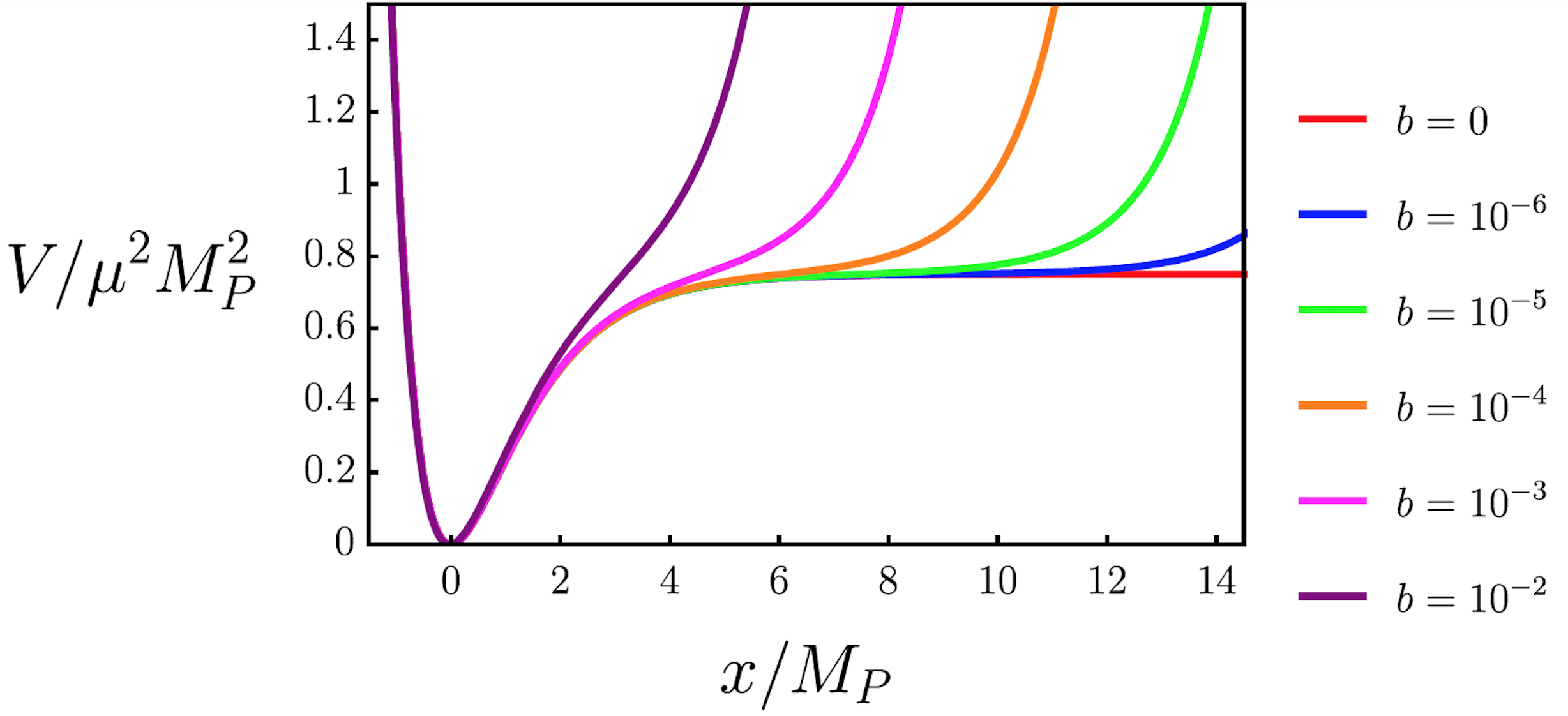}
	\caption{Potential for different values of $b$. For $b=0$ the exact Starobinsky limit is obtained and the potential has a flat plateau. For $b\sim 10^{-5}$ the potential retains a quite flat region, while for $b\gtrsim 10^{-4}$ the potential loses its flatness.}
	\label{fig:potential}
\end{figure}

We begin by computing the supergravity scalar potential from Eqs.~\ref{eq:ka_function} and \ref{eq:scalar_potential},
in the hidden sector using $\widetilde{W}_{hid}$~\ref{eq:W_inflation_limit} and $\widetilde{K}_{hid}$~\ref{eq:ka_potential_y_variables} only. We follow the same treatment as in~\cite{Ellis:2013xoa} and assume that the $T$ field is fixed with a vacuum expectation value of $\vev{Re T}=c/2$ and $\vev{Im T}=0$, corresponding to $\vev{y_2}=0$\footnote{The modulus field $y_2$ is stabilized when including a term $\propto y_2^4$ as in Eq.~\ref{eq:ka_stabilizing_term}, for a better understanding see~\cite{Ellis:2013nxa} and Section~\ref{sec:stabilizing}.}.  The minimum of the potential is always given by $V=0$, for both cases I and II, $V=0$ is found for
\begin{equation}
	y_1=\frac{\sqrt{3}}{2} \left(1\pm\sqrt{1+4b}\right) M_P,
	\label{eq:y1minimum}
\end{equation}
such that supersymmetry is spontaneously broken in the hidden sector. For $b=0$, $y_1=0$ only if we have the minus sign and we restrict ourselves to this case in the following. This model reproduces the effective potential of the Starobinsky model for $\mathcal{R}e$ $y_1$. The dynamical field $y_1$ can be converted into a canonically-normalized inflaton field $x$ by the transformation~\cite{Ellis:2013xoa, Ellis:2013nxa}
\begin{equation}
	y_1=\pm \sqrt{3}M_P\tanh\left(\frac{\chi}{\sqrt{3}M_P}\right)=\pm \sqrt{3}M_P\tanh\left(\frac{x}{\sqrt{6}M_P}\right),
	\label{eq:canonically_normalized}
\end{equation}
where $\chi=(x+iy)/\sqrt{2}$ and the latter equality holds for $y=0$. The imaginary part of the inflaton is fixed to $y=0$ by the potential~\cite{Ellis:2013xoa,Romao:2017uwa} since the potential is always minimized by $y=0$ in the range of interest of the inflaton field $x$. We use the positive sign in Eq.~\ref{eq:canonically_normalized} and write the potential in terms of the inflaton field $x$
\begin{equation}
V=\frac{3}{4}\mu^2M_P^2\left[1+b+(b-1)\cosh\left(\sqrt{\frac{2}{3}}\frac{x}{M_P}\right)+\sinh\left(\sqrt{\frac{2}{3}}\frac{x}{M_P}\right)\right]^2,
\end{equation}
where we have fixed $\vev{y_2}=0$. In terms of the inflaton field $x$, the minimum of the potential is found for
\begin{equation}
x_0=\sqrt{6} M_P \tanh^{-1}\left(\frac{1}{2}(1- \sqrt{1+4b})\right). 
\end{equation}
The exact Starobinsky limit is realized for $b=0$\footnote{For $b=0$, the scalar potential can be written as $V=\frac{3}{4}\mu^2M_P^2\left(1-e^{-\sqrt{2/3}x/M_P}\right)^2$, which is exactly the Starobinsky potential~\cite{R2}.}, while small values for $b$ represents a small deviation from the Starobinsky limit as shown in Fig.~\ref{fig:potential}.

 As $b$ deviates from zero, the value of the field at the global minimum $x_0$ shifts away from zero, while maintaining $V=0$. Although the scalar potential vanishes, the superpotential $W$ and the \ka{1} function $G$ is different from zero, generating explicit soft supersymmetry-breaking terms of the required form in the effective low-energy  Lagrangian. The gravitino becomes massive, see Eq.~\ref{eq:gravitino_mass}, and sets the overall scale of the soft parameters 
\begin{equation}
m_{3/2}=- \frac{3}{2}\mu b^2 +\mathcal{O}(b^3),
\end{equation}
where we have expanded at lowest order in $b$. The exact analytical expression can be found in Appendix~\ref{app:Soft_parameters}.

\begin{figure}[t!]
	\centering
\includegraphics[scale=0.6]{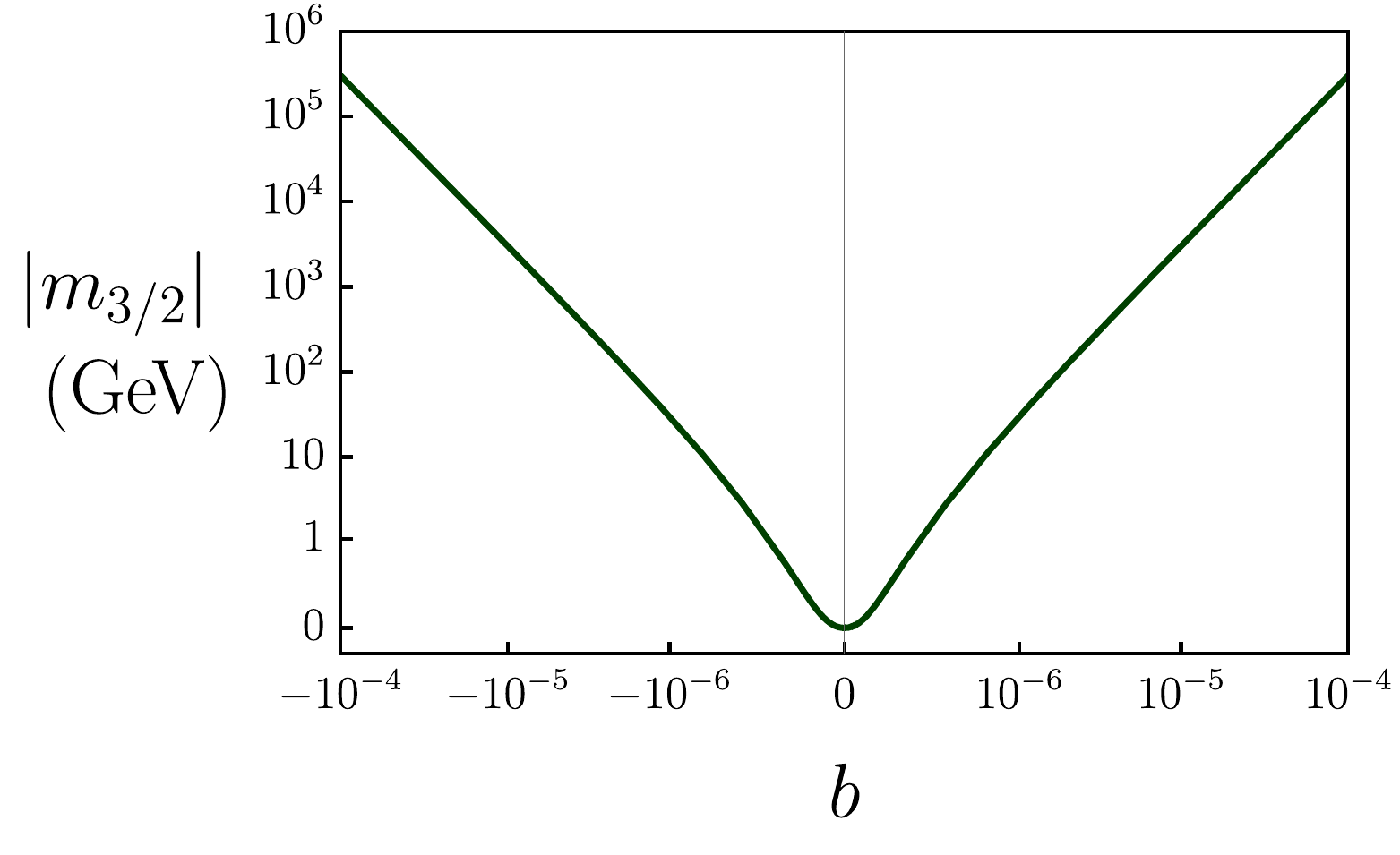}
\caption{Gravitino mass as a function of the parameter $b$ accounting for the Polonyi term. As shown in the previous 
Figure~\ref{fig:potential} (for details see \cite{Romao:2017uwa}), successful inflation requires $b\lesssim 10^{-4}$,
corresponding here to a strict upper bound for the gravitino mass $m_{3/2}< 10^3$ TeV.}
\label{fig:gravitino_mass}
\end{figure}

As explained in \cite{Romao:2017uwa},
the limits from inflation suggest that the parameters  $\mu\simeq10^{-5}M_P\simeq2 \times 10^{13}$ GeV and $b\lesssim10^{-4}$. 
For fixed $x_{*}=5.35 M_P$ (where $x_{*}$ is the value of the field when inflation starts) 
we see that there is an approximate quadratic dependence of the gravitino mass on the parameter $b$, as shown in Fig.~\ref{fig:gravitino_mass},
where we have rescaled the results around the origin as in \cite{Romao:2017uwa}.
For $b=0$, we see that $m_{3/2}=0$, so supersymmetry is unbroken and we are left with the Wess-Zumino superpotential limit leading to Starobinsky inflation. However, from Fig.~\ref{fig:potential}, we see that for small values of $b\lesssim10^{-4}$, the potential retains a plateau where inflation will happen and the corresponding limit on the gravitino mass may be read off from 
Fig.~\ref{fig:gravitino_mass} as $m_{3/2} < 10^{3}$ TeV.

\subsubsection{Stabilizing the modulus field}
\label{sec:stabilizing}
 We introduced a term in the \ka{1} potential $\propto y^4_2/\Lambda^2$ to assure the stabilization of the field $y_2$ during inflation, see Eq.~\ref{eq:ka_stabilizing_term}. Additionally, we can compute the mass of the modulus field $y_2$, $m_{y_2}$, and the mass of the field $y_1$, $m_{y_1}$, during inflation. As a benchmark point, we choose $x=5M_P$ and $\Lambda=0.01M_P$, and we find that the masses are $m_{y_1}\sim 10^{13}$ GeV and $m_{y_2}\sim 10^{17} \text{ GeV}$. The fact that $m_{y_2}\gg m_{y_1}$ is valid, not only for this benchmark point, but during the whole inflationary trajectory for $\Lambda\lesssim 0.1 M_P$, means that the single field approximation is justified during inflation.
 
At the end of inflation, when $y_1$ is at its minimum (see Eq.~\ref{eq:y1minimum}), the masses of the fields $y_1$ and $y_2$ are given by
\begin{equation}
	m_{y_1}=\sqrt{2}\mu (1-b) +\mathcal{O}(b^2), \quad m_{y_2}=\sqrt{2}\mu b +\mathcal{O}(b^2).
\end{equation}
\begin{figure}
	\centering
\includegraphics[scale=0.2]{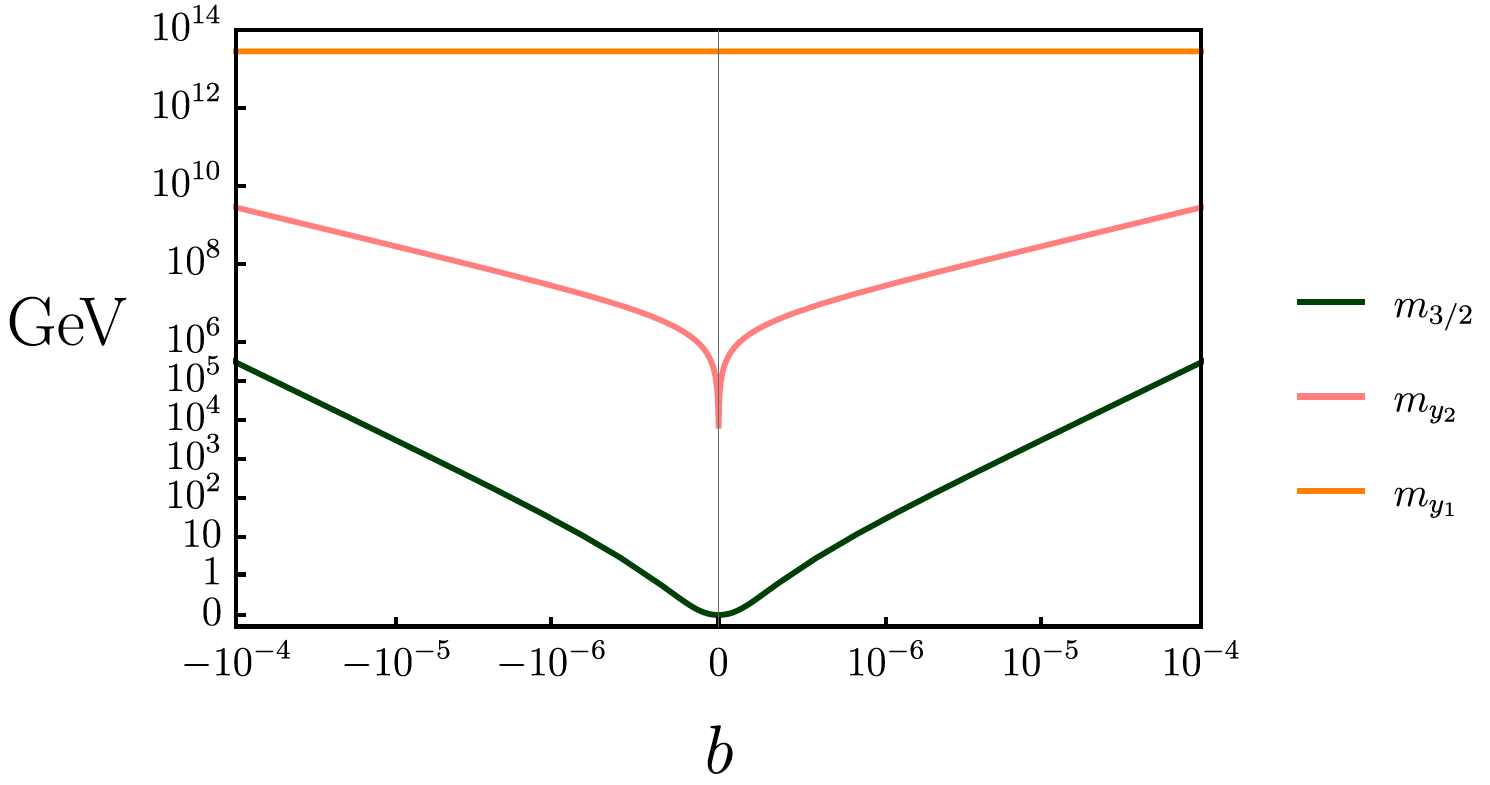}
\caption{Masses of the gravitino, the modulus field $y_2$ and the field $y_1$, for different values of $b$, at the end of inflation. The masses are evaluated when $y_1$ is at its minimum, given by Eq.~\ref{eq:y1minimum}. The modulus field is strongly stabilized with a mass $m_{y_2}\gg m_{3/2}$.}
\label{fig:y1_y2_mass}
\end{figure}Fig.\ref{fig:y1_y2_mass} shows the gravitino mass as well as the mass of the modulus field $y_2$ and the mass of the field $y_1$. The modulus $y_2$ is strongly stabilized at the end of inflation, where $m_{y_2} \gg m_{3/2}$. 
\subsection{Visible sector potential and SUSY breaking}
We now compute the supergravity scalar potential from Eqs.~\ref{eq:ka_function} and \ref{eq:scalar_potential},
including the visible sector superfields, 
using either Eq.\ref{eq:Ka_SM_inside} (case I) or Eq.\ref{eq:Ka_SM_outside} (case II),
with the total superpotential in Eq.\ref{W}.
With the addition of the visible sector superpotential in Eq.~\ref{eq:SM_superpotential}, we are now able to compute the soft supersymmetry breaking mass-squared, bilinear, and trilinear parameters, $m_0, B_0 \text{ and } A_0$ respectively. 

\subsubsection{Case I: no-scale SUSY breaking}
\begin{figure}[h!]
	\centering
	\begin{subfigure}{0.45\textwidth}
		\includegraphics[scale=0.5]{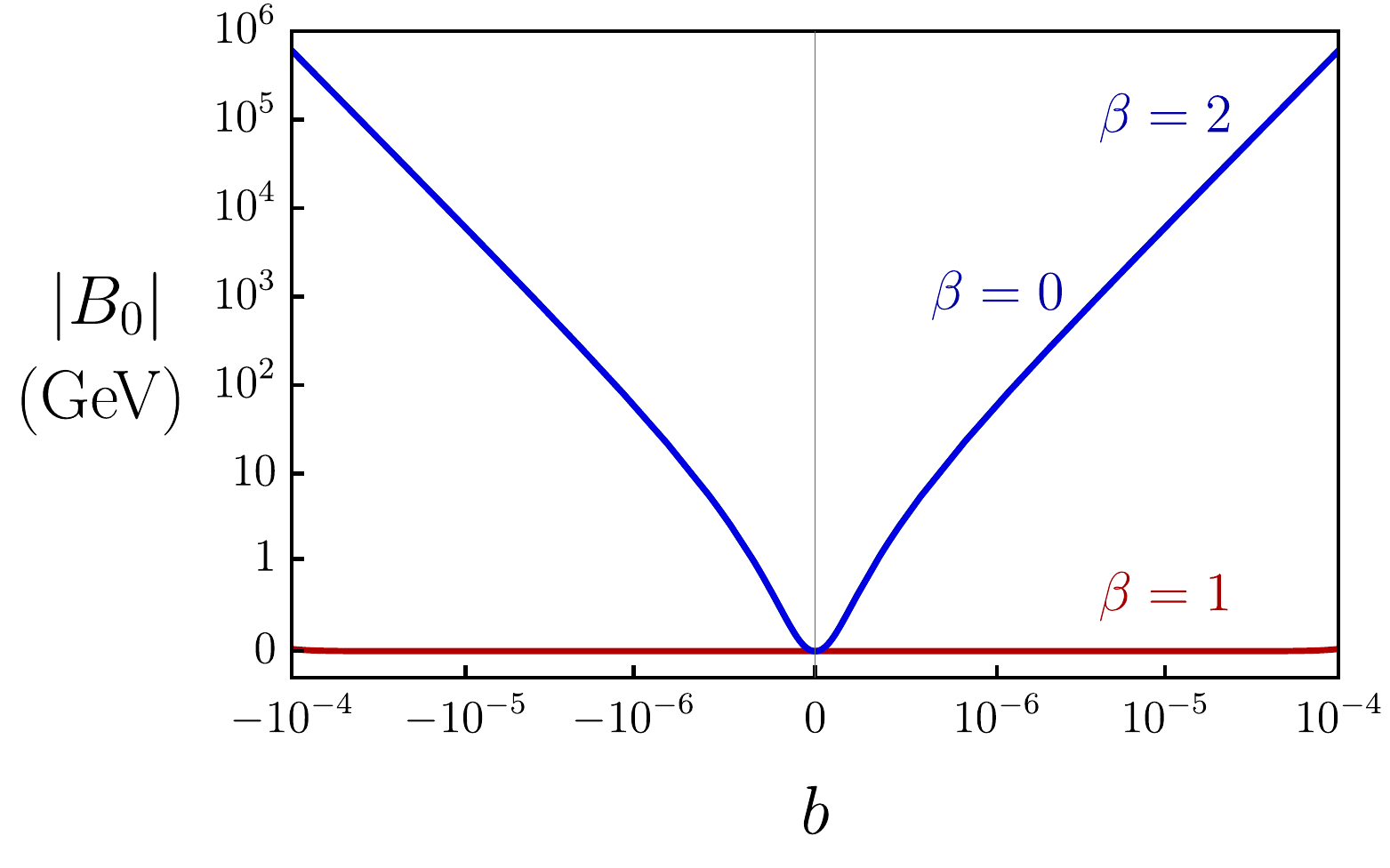}
		\caption{The bilinear soft supersymmetry-breaking parameter.}
		\label{fig:B0_soft_breaking_parameters}
	\end{subfigure}%
	\ \ \ \ \
	\begin{subfigure}{0.45\textwidth}
		\includegraphics[scale=0.5]{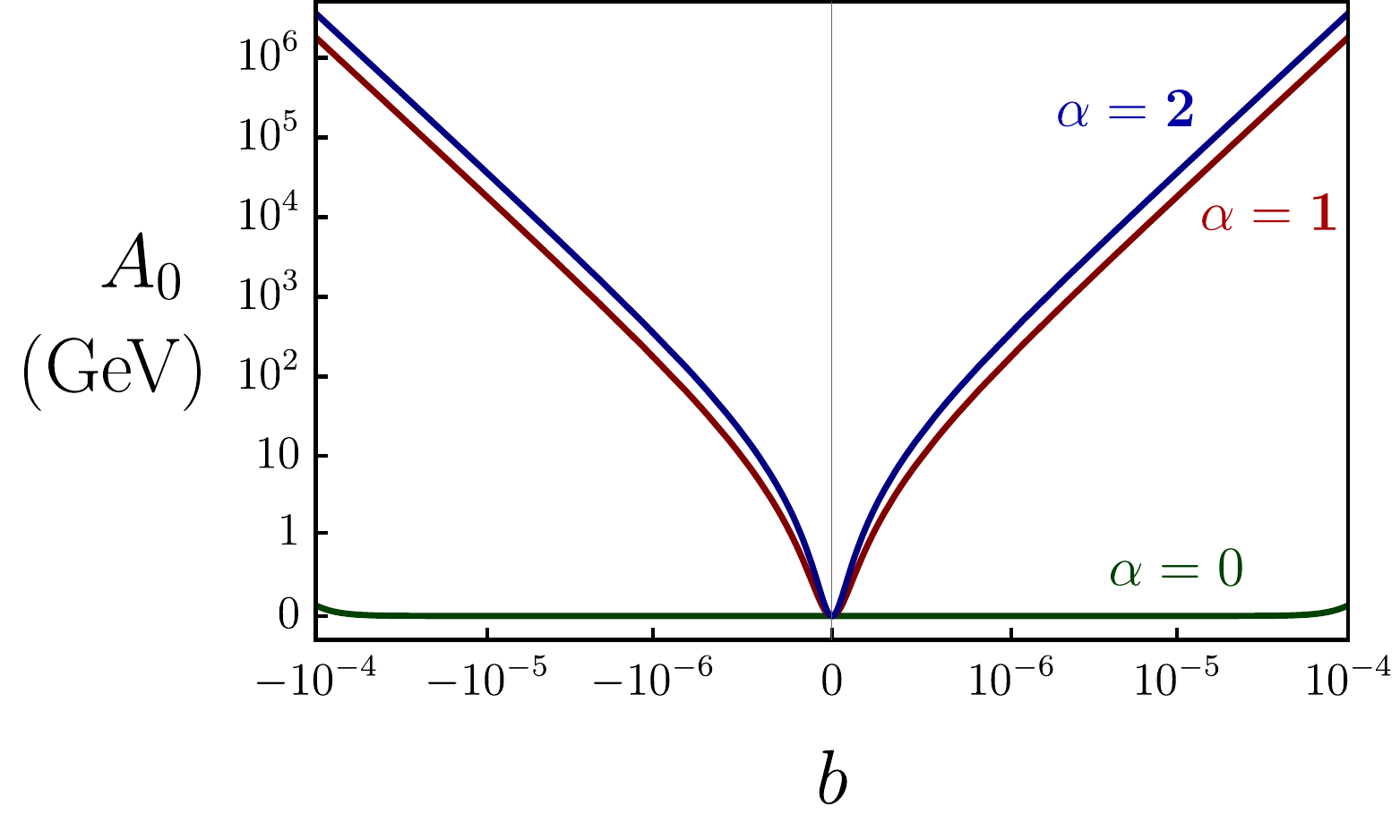}
		\caption{The trilinear soft supersymmetry-breaking parameter.}
		\label{fig:A0_soft_breaking_parameters}
	\end{subfigure}
	\caption{Soft supersymmetry-breaking parameters as a function of the parameter $b$, accounting for the Polonyi term, for different modular weights $\alpha$ and $\beta$.  }
	\label{fig:soft_breaking_parameters}
\end{figure}

For the Case I, where the SM superfields are inside the Log of the \ka{1} potential, the soft supersymmetry breaking parameters become
\begin{equation}
\begin{split}
&m_0=0, \\
&\frac{A_0}{m_{3/2}}=-6\alpha -3\left(4+\alpha\right)b^2 +\mathcal{O}(b^3), \qquad \text{ Case I } \\
&\frac{B_0}{m_{3/2}}=2(1-\beta) -\left(1+\beta\right)b^2 +\mathcal{O}(b^3). 
\end{split}
\label{eq:soft_parameters}
\end{equation}
The prediction for $m_0=0$ is the familiar result of no-scale SUSY breaking.
The exact analytical functions are given in Appendix~\ref{app:Soft_parameters}, while the expressions in Eq.~\ref{eq:soft_parameters} are found when expanding in powers of $b$ and hold for our range of interest 
$b\lesssim 10^{-4}$. The mass-squared term $m_0$ is zero while the bilinear and trilinear parameters will depend on the choice of the modular weights $\alpha$ and $\beta$, as shown in Fig.~\ref{fig:soft_breaking_parameters}.  
The main effect of switching on $b$ is to increase the gravitino mass, since the terms proportional to $b^2$ are negligible.
The special choice of $\alpha=0$ and $\beta=1$, corresponds to the pure no-scale option where $m_0=B_0=A_0=0$ for $b\lesssim10^{-4}$, in which case the supersymmetry breaking in the low scale energy can be produced via a non-minimal gauge kinetic term generating non-zero gaugino masses $M_{1/2}\neq0$.

For the rest of cases, where $\alpha\neq0$ and $\beta\neq1$, we can set some limits in the bilinear and trilinear couplings due to the constraint on the parameter $b\lesssim 10^{-4}$ coming from inflation such that $B_0 < 10^{3}$ TeV and $A_0\lesssim 10^{3}$ TeV. We assume the condition $m_{3/2}\gg M_{1/2}$ to avoid constraints from Big Bang nucleosynthesis on gravitino decays.  In this case, we cannot find a viable phenomenological region due to the relation between the trilinear coupling and the gravitino mass $A_0\simeq-6\alpha m_{3/2}$ in Eq.~\ref{eq:soft_parameters}, implying $A_0\gg M_{1/2}$ which leads to tachyonic sfermion masses. Furthermore, for $\beta\neq1$, there is no value of $\tan \beta$ which minimizes the Higgs potential and which could give the correct Higgs Mass $m_H=125\pm 0.24$ GeV. Therefore, the only case we can further explore is the case $\alpha=0$ and $\beta=1$, corresponding to $m_0=A_0=B_0=0$.

It has been proven in~\cite{Ellis:2010jb} the no-scale boundary conditions $m_0=A_0=B_0=0$ when becoming universal at some unification scale $M_{in}$ above the GUT scale, are compatible with low-energy constraints. They show a scenario based on $SU(5)$ model where the superpotential contains terms $W \ni \lambda H \Sigma \bar{H}+(\lambda'/6)\text{Tr}\Sigma^3$, where $H, \overline{H}$ and $\Sigma$ are $\mathbf{5}, \mathbf{\bar{5}}$ and $\mathbf{24}$ Higgs representations. For different values of $\lambda$ and $\lambda'$, the region of $M_{in}$, $M_{1/2}$ and $\tan \beta$ has been studied, taking into account phenomenological constraints on supersymmetric particles and the cosmological LSP density.  Although, the LHC has imposed additional constraints, via the measurement of the Higgs Mass and the decay $B_s\rightarrow\mu^+\mu^-$, the model is still consistent with the LHC data~\cite{Ellis:2013nka, Ellis:2015xna}. For example, for the values $\lambda=-0.1 $ and $\lambda'=2$, the region with $M_{1/2} \in (1000,1500)$ GeV and $M_{in} \in (10^{17},10^{18})$ GeV is consistent with the relic cold dark matter density, a Higgs mass of $m_H\sim125$ GeV and the experimental measurements of $B_s\rightarrow\mu^+ \mu^-$.

\subsubsection{Case II: pure gravity SUSY breaking}

As an alternative to the no-scale possibility, we can have the Standard Model superfields $y_{vis}$ outside the logarithm.
This is called Case II in Eq.~\ref{eq:Ka_SM_outside}. For small $b$, where inflation works, we find that the soft supersymmetry breaking parameters are
\begin{equation}
\begin{split}
&m_0=m_{3/2}, \\
&\frac{A_0}{m_{3/2}}=-6\alpha -3\left(4+\alpha\right)b^2 +\mathcal{O}(b^3), \qquad \text{ Case II } \\
&\frac{B_0}{m_{3/2}}=2(1-\beta) -\left(1+\beta\right)b^2 +\mathcal{O}(b^3),
\end{split}
\label{eq:soft_parameters_caseII}
\end{equation}  where the relations for $A_0$ and $B_0$ are kept the same as in  Eq.~\ref{eq:soft_parameters} while the universal soft scalar mass term $m_0$ is now different from zero and equal to the gravitino mass $m_{3/2}$. 
Note that the soft scalar mass is equal to the gravitino mass,
while the soft gaugino mass will be assumed to be much smaller, as in anomaly mediated SUSY breaking.
As before, the main effect of switching on $b$ is to increase the gravitino mass, since the terms proportional to $b^2$ are negligible.

For the same reasons as before, the general cases $\alpha\neq0$ and $\beta \neq 1$ are not viable phenomenological choices. 
As mentioned, we impose $m_{3/2}\gg M_{1/2}$, meaning $A_0 \gg M_{1/2}$ (see Eq.~\ref{eq:soft_parameters_caseII}). This will lead to tachyonic sfermion masses. Similarly, for $\beta\neq 1$ we cannot find a value of $\tan \beta$ which reproduces the correct Higgs Mass. Therefore, we will focus in the case $\alpha=0$ and $\beta=1$  on the following, corresponding to $m_0=m_{3/2}$ and $A_0=B_0=0$.

The special case $\alpha=0$ and $\beta=1$, meaning $A_0\approx B_0\approx 0$, is called pure gravity-mediated (PGM) models~\cite{Ibe:2007, Dudas:2012wi,Evans:2013lpa, Evans:2013dza, Evans:2014xpa, Evans:2014pxa, Evans:2015bxa} where the gaugino masses, $A$ and $B$ terms are determined by anomaly mediation~\cite{Dine:1992} leaving only the gravitino mass $m_{3/2}=m_0$ as a free parameter.  
Note that in this case since $b\lesssim 10^{-4}$, Eq.\ref{eq:soft_parameters_caseII} gives $A_0=B_0=0$ to excellent approximation.
The main phenomenological effect of the Polonyi term is then to set a limit on the gravitino mass from the requirement of successful inflation.
These PGM models with GUT scale universality~\cite{Dudas:2012wi} are phenomenologically viable when including a Giudice-Masiero term in the \ka{1} potential~\cite{Giudice:1988yz}, which allows to choose $\tan \beta$ as a second free parameter. For successful electroweak symmetry breaking (EWSB), $\tan \beta$ is restricted to a narrow range from about $1.7-2.5$~\cite{Evans:2013lpa, Evans:2015bxa}. Then, the Higgs mass $m_H\sim125$ GeV is obtained when the gravitino mass is in the range $300-1500$ TeV. In our case we have the additional constraint from inflation requiring $m_{3/2}<1000$ TeV, see Fig.~\ref{fig:gravitino_mass}. Therefore, the Case II, in which $m_0=m_{3/2}$, for $\alpha=0$ and $\beta=1$ has an allowed parameter space when including the Giudice-Masiero term which is also compatible with inflation and corresponds to $b\sim 10^{-4}$.

Recently, PGM models have been studied in an SU(5) GUT model~\cite{Evans:2019oyw},
allowing two new parameters, a high energy scale above the GUT scale, $M_{in}$ and a new coupling $\lambda \bar{H} \Sigma H $,
where $\Sigma$ is an $SU(5)$ adjoint Higgs .
They find a viable parameter space as long as the Higgs coupling $\lambda$ is relatively large,
for a relaxed $\tan \beta$, away from $\tan \beta\sim 2$, opening up the parameter space considerably.
In the case where higher dimensional operators involving are included, proton decay can be within reach of future experiments. Moreover, in some regions of parameter space the bino can be degenerate with the wino or gluino, giving an acceptable dark matter relic density.

\section{Conclusion}
\label{conclusion}

In this paper 
we have studied a version of Starobinsky-like inflation in no-scale SUGRA where a Polonyi term in the hidden sector
breaks SUSY after inflation, providing a link between 
the gravitino mass and inflation. 
The linear Polonyi term provides a simple way to break SUSY after inflation, with the requirement of 
successful inflation leading to an upper bound on the gravitino mass $m_{3/2}<10^3$ TeV, with gaugino masses 
considerably less than this.

The progress made in this paper is to extend the previous hidden sector Polonyi
theory (proposed by one of us) to include the visible sector and to calculate the resulting
soft-SUSY breaking parameters in terms of the modular weights $\alpha$ and $\beta$ for the bilinear and trilinear 
parts of the superpotential and a choice of \ka{1} potential. 
In this case, we can write the soft-SUSY breaking parameters in terms of $\alpha$, $\beta$ and the parameter $b$ accounting for the Polonyi term. 
In case I we included the MSSM matter and Higgs 
superfields inside the Log of the \ka{1} potential following a no-scale approach. In the second case, Case II, these superfields are added via minimal kinetic terms. 

We were thereby led to familiar phenomenological possibilities for supersymmetry (SUSY) breaking, 
based on no-scale SUSY breaking and pure gravity mediated SUSY breaking, but with the strict bound coming 
from our Polonyi model that $m_{3/2}<10^3$ TeV.
However we found that the only phenomenological viable choices for both cases I and II, are the ones with $\alpha=0$ and $\beta=1$, corresponding to $A_0=B_0=0$. 

In case I, in addition to $A_0=B_0=0$, the scalar soft mass term is zero $m_0=0$, as in the 
no-scale SUSY breaking approach, in which case SUSY breaking at low energies is produced via non-zero gaugino masses $M_{1/2}\neq0$. Models like this one had been discussed by ENO and have been shown to be compatible with LHC data,
with all superpartners potentially observable at  LHC or the FCC.

In case II, in addition to $A_0=B_0=0$, the scalar mass is equal to the gravitino mass $m_0=m_{3/2}$,
with $M_{1/2}\ll m_{3/2}$ as in anomaly mediation models, with such a scenario referred to as PGM.
Such PGM models are viable models after the inclusion of a Giudice-Masiero term in the \ka{1} potential, for a large gravitino mass of order 100-1000 TeV, compatible with our inflation bound,
and can evade LHC searches while still providing a good dark matter candidate and gauge coupling unification,
with squarks and sleptons very heavy, while the 
gauginos remain light and observable at the LHC or the FCC.

In conclusion, Starobinsky-like inflation in no-scale SUGRA models with a Polonyi term
provides a promising setting for both 
inflation and SUSY breaking
within a well motivated particle physics framework.
The Polonyi term provides a link between 
the gravitino mass and inflation leading to a strict upper bound on the gravitino mass $m_{3/2}<10^3$ TeV.
We have seen that under reasonable assumptions, the soft-SUSY breaking parameters can be calculated,
leading to either no-scale SUGRA or PGM patterns, where the 
results presented here could form the basis of future phenomenological studies.
In particular, since gauginos are significantly lighter than $m_{3/2}$, 
this suggests that SUSY could be discovered at the LHC or FCC.

\section*{Acknowledgments}
SFK thanks the CERN Theory group for hospitality and 
acknowledges the STFC Consolidated Grant ST/L000296/1.
SFK and EP acknowledge
the European Union's Horizon 2020 Research and Innovation programme under Marie Sk\l{}odowska-Curie grant agreements Elusives ITN No.\ 674896 and InvisiblesPlus RISE No.\ 690575.

\appendix
\section{Soft-supersymmetry breaking parameters}
\label{app:Soft_parameters}
We present the analytic results for the gravitino mass and the soft-supersymmetry breaking parameters as a function of the modular weights $\alpha$ and $\beta$ and the parameter $b$, which takes into account the contribution from the Polonyi term in the superpotential. For both cases I and II, in which we include the Standard Model superfields inside or outside the logarithm in the \ka{1} potential, see Eqs.~\ref{eq:Ka_SM_inside} and~\ref{eq:Ka_SM_outside}, the bilinear, $B_0$ and trilinear $A_0$ parameters are the same and the only difference is in the soft scalar mass term $m_0$, vanishing in case I while being equal to the gravitino mass in case II.
\begin{equation}
\begin{split}
m_{3/2}=&-\mu\frac{-1+\sqrt{1+4b}+b\left(-6+4\sqrt{1+4b}\right)}{\sqrt{2}\left(1-2b+\sqrt{1+4b}\right)^{3/2}} \\ \\
m_0=&0 \text{ (Case I)}\quad  \text{or} \quad m_0=m_{3/2} \text{ (Case II) } , \\ \\
A_0 =&-\mu \frac{3 \left(\sqrt{4 b+1}-1\right)^2 \left(-8 b+\sqrt{4 b+1}-1\right)}{\sqrt{2} \left(-2 b+\sqrt{4 b+1}+1\right)^3} \\
&\frac{ \left(b \left(\alpha -8 (\alpha -2) b+5 \alpha  \sqrt{4 b+1}-10 \sqrt{4 b+1}+14\right)-2 \sqrt{4 b+1}+2\right)}{ \sqrt{\left(-\sqrt{4 b+1}+2 b \left(-5 \sqrt{4 b+1}+b \left(16 b-12 \sqrt{4 b+1}+21\right)+6\right)+1\right) }}, \\ \\
B_0 =&- \mu \frac{\left(\sqrt{4 b+1}-1\right)^2 \left(-8 b+\sqrt{4 b+1}-1\right)}{\left(-2 b+\sqrt{4 b+1}+1\right)^3}\\
&\frac{ \left(b \left(-8 b (\beta -2)+5 \beta  \sqrt{4 b+1}+\beta -10 \sqrt{4 b+1}+6\right)-\sqrt{4 b+1}+1\right)}{\sqrt{-2 \sqrt{4 b+1}+4 b \left(-5 \sqrt{4 b+1}+b \left(16 b-12 \sqrt{4 b+1}+21\right)+6\right)+2}}.
\end{split}
\label{eq:soft_parameters_appendix}
\end{equation}


\begin{thebibliography}{99}
	\setlength{\itemsep}{0em}
	
			
		\bibitem{Guth:1980zm}
	Alan~H. Guth.
	\newblock {\em Phys. Rev.}, D23:347--356, 1981.
	
		\bibitem{Linde:1981mu}
	Andrei~D. Linde.
	\newblock {\em Phys. Lett.}, B108:389--393, 1982.
	
	\bibitem{Mukhanov:1981xt}
	Viatcheslav~F. Mukhanov and G.~V. Chibisov.
	\newblock {\em JETP Lett.}, 33:532--535, 1981.
	\newblock [Pisma Zh. Eksp. Teor. Fiz.33,549(1981)].

\bibitem{Albrecht:1982wi}
	Andreas Albrecht and Paul~J. Steinhardt.
	\newblock {\em Phys. Rev. Lett.}, 48:1220--1223, 1982.

	
	\bibitem{Linde:1983gd}
	Andrei~D. Linde.
	\newblock {\em Phys. Lett.}, B129:177--181, 1983.
	
	\bibitem{Linde:2007fr}
	A.~D.~Linde,
	Lect.\ Notes Phys.\  {\bf 738} (2008) 1
	doi:$0.1007/978-3-540-74353-8_1$
	[arXiv:0705.0164 [hep-th]].
	
	 \bibitem{Linde1990}
	Andrei~D. Linde.
	\newblock {\em Contemp. Concepts Phys.}, 5:1--362, 1990.
	
		
	\bibitem{LythRiotto1999}
	David~H. Lyth and Antonio Riotto.
	\newblock {\em Phys. Rept.}, 314:1--146, 1999.

	
	\bibitem{Ade:2015lrj}
	P.~A.~R. Ade et~al.
	\newblock {\em Astron. Astrophys.}, 594:A20, 2016.
	
	  \bibitem{MartinRingevalVennin2014}
	Jerome Martin, Christophe Ringeval, and Vincent Vennin.
	\newblock {\em Phys. Dark Univ.}, 5-6:75--235, 2014.


	
	\bibitem{R2}
A.~A.~Starobinsky,
  Phys.\ Lett.\ B {\bf 91}, 99 (1980);
  A.~A.~Starobinsky,
  Sov.\ Astron.\ Lett.\  {\bf 9}, 302 (1983).
  
    \bibitem{HI}
  F.~Bezrukov and M.~Shaposhnikov,
 JHEP {\bf 0907}, 089 (2009)
 [arXiv:0904.1537 [hep-ph]].
 
 
 
 \bibitem{others}
A.~Linde, M.~Noorbala and A.~Westphal,
 JCAP {\bf 1103}, 013 (2011)
 [arXiv:1101.2652 [hep-th]];
  S.~Ferrara, R.~Kallosh, A.~Linde, A.~Marrani and A.~Van Proeyen,
 Phys.\ Rev.\ D {\bf 83} (2011) 025008
 [arXiv:1008.2942 [hep-th]].
 
\bibitem{Copeland:1994vg}
  E.~J.~Copeland, A.~R.~Liddle, D.~H.~Lyth, E.~D.~Stewart and D.~Wands,
  Phys.\ Rev.\ D {\bf 49} (1994) 6410
  doi:10.1103/PhysRevD.49.6410
  [astro-ph/9401011];
  G.~R.~Dvali, Q.~Shafi and R.~K.~Schaefer,
  Phys.\ Rev.\ Lett.\  {\bf 73} (1994) 1886
  doi:10.1103/PhysRevLett.73.1886
  [hep-ph/9406319].
  
    

  
    
	\bibitem{Ellis:1982ed}
John~R. Ellis, Dimitri~V. Nanopoulos, Keith~A. Olive, and K.~Tamvakis.
\newblock {\em Phys. Lett.}, B118:335, 1982.

\bibitem{Ellis:1982dg}
John~R. Ellis, Dimitri~V. Nanopoulos, Keith~A. Olive, and K.~Tamvakis.
\newblock {\em Phys. Lett.}, B120:331--334, 1983.


  
  	
  \bibitem{Ellis:1982ws}
  John~R. Ellis, Dimitri~V. Nanopoulos, Keith~A. Olive, and K.~Tamvakis.
  \newblock {\em Nucl. Phys.}, B221:524--548, 1983.
	
	
\bibitem{Lyth:1984yz}
  D.~H.~Lyth,
  Phys.\ Lett.\  {\bf 147B} (1984) 403
   Erratum: [Phys.\ Lett.\  {\bf 150B} (1985) 465].
  doi:10.1016/0370-2693(84)91391-1
	
		\bibitem{YanagidaSUGRA}
	F.~Bjorkeroth, S.~F.~King, K.~Schmitz and T.~T.~Yanagida,
	arXiv:1608.04911 [hep-ph].
	K.~Nakayama, F.~Takahashi and T.~T.~Yanagida,
	Phys.\ Lett.\ B {\bf 757} (2016) 32
	doi:10.1016/j.physletb.2016.03.051
	[arXiv:1601.00192 [hep-ph]].
	K.~Harigaya, M.~Kawasaki and T.~T.~Yanagida,
	Phys.\ Lett.\ B {\bf 741} (2015) 267
	doi:10.1016/j.physletb.2014.12.053
	[arXiv:1410.7163 [hep-ph]].
	S.~Hellerman, J.~Kehayias and T.~T.~Yanagida,
	Phys.\ Lett.\ B {\bf 742} (2015) 390
	doi:10.1016/j.physletb.2015.02.019
	[arXiv:1411.3720 [hep-ph]].
	K.~Schmitz and T.~T.~Yanagida,
	Phys.\ Rev.\ D {\bf 94} (2016) no.7,  074021
	doi:10.1103/PhysRevD.94.074021
	[arXiv:1604.04911 [hep-ph]].
	J. R. Ellis,
	M. Raidal and T. Yanagida, Phys. Lett. B 581, 9 (2004) [hep-ph/0303242]. K. Nakayama, F. Takahashi and T. T. Yanagida, Phys. Lett. B
	730, 24 (2014) [arXiv:1311.4253 [hep-ph]].
	
	
	
	
	
	
	
	
	\bibitem{AntushcSUGRAinflation}
	S. Antusch, M. Bastero-Gil, S. F. King
	and Q. Shafi, Phys. Rev. D 71, 083519 (2005) [hep-ph/0411298].
	S.~Antusch and D.~Nolde,
	JCAP {\bf 1509} (2015) no.09,  055
	doi:10.1088/1475-7516/2015/09/055
	[arXiv:1505.06910 [hep-ph]].
	S.~Antusch and K.~Dutta,
	Phys.\ Rev.\ D {\bf 92} (2015) 083503
	doi:10.1103/PhysRevD.92.083503
	[arXiv:1505.04022 [hep-ph]].
	
	\bibitem{KalloshSUGRAinflation}
	R. Kallosh, A. Linde, D. Roest and T. Wrase, arXiv:1607.08854 [hep-th]. M. K. Gaillard, H. Murayama and K. A. Olive, Phys. Lett.
	B 355, 71 (1995) [hep-ph/9504307]. K. Kadota and J. Yokoyama, Phys. Rev. D
	73, 043507 (2006) [hep-ph/0512221]. H. Murayama, K. Nakayama, F. Takahashi
	and T. T. Yanagida, Phys. Lett. B 738, 196 (2014) [arXiv:1404.3857 [hep-ph]].
	K. Nakayama, F. Takahashi and T. T. Yanagida, JCAP 1308, 038 (2013) [arXiv:1305.5099 [hep-ph]].
	Phys. Lett. B 737, 151 (2014) [arXiv:1407.7082 [hep-ph]]. J. L. Evans, T. Gherghetta
	and M. Peloso, Phys. Rev. D 92, no. 2, 021303 (2015) [arXiv:1501.06560 [hep-ph]].
	A. K. Saha and A. Sil, JHEP 1511, 118 (2015) [arXiv:1509.00218 [hep-ph]].
	
	
	\bibitem{Ellis:1984bf}
	J.~R.~Ellis, K.~Enqvist, D.~V.~Nanopoulos, K.~A.~Olive and M.~Srednicki,
	Phys.\ Lett.\  {\bf 152B} (1985) 175
	Erratum: [Phys.\ Lett.\  {\bf 156B} (1985) 452].
	doi:10.1016/0370-2693(85)91164-5
	
		\bibitem{BG}
	P.~Binetruy and M.~K.~Gaillard,
	Phys.\ Lett.\ B {\bf 195} (1987) 382;
	H.~Murayama, H.~Suzuki, T.~Yanagida and J.~Yokoyama,
	Phys.\ Rev.\  D {\bf 50}, 2356 (1994)
	[arXiv:hep-ph/9311326];
	S.~Antusch, M.~Bastero-Gil, K.~Dutta, S.~F.~King and P.~M.~Kostka,
	Phys.\ Lett.\ B {\bf 679} (2009) 428
	[arXiv:0905.0905 [hep-th]].
	
	\bibitem{HG}
	S.~Antusch, K.~Dutta, J.~Erdmenger and S.~Halter,
	JHEP {\bf 1104} (2011) 065
	doi:10.1007/JHEP04(2011)065
	[arXiv:1102.0093 [hep-th]];
	S.~Antusch and F.~Cefalà,
	JCAP {\bf 1310} (2013) 055
	doi:10.1088/1475-7516/2013/10/055
	[arXiv:1306.6825 [hep-ph]].
	
	
	
	\bibitem{Y}
	M.~Kawasaki, M.~Yamaguchi and T.~Yanagida,
	Phys.\ Rev.\ Lett.\  {\bf 85} (2000) 3572
	[hep-ph/0004243];
	K.~Nakayama, F.~Takahashi and T.~T.~Yanagida,
	arXiv:1303.7315 [hep-ph].
	
	
	\bibitem{Davis:2008fv}
	S.~C.~Davis and M.~Postma,
	JCAP {\bf 0803}, 015 (2008)
	[arXiv:0801.4696 [hep-ph]].
	
	\bibitem{klor}
	R.~Kallosh and A.~Linde,
	JCAP {\bf 1011}, 011 (2010)
	[arXiv:1008.3375 [hep-th]];
	R.~Kallosh, A.~Linde and T.~Rube,
	Phys.\ Rev.\  D {\bf 83}, 043507 (2011)
	[arXiv:1011.5945 [hep-th]];
	R.~Kallosh, A.~Linde, K.~A.~Olive and T.~Rube,
	Phys.\ Rev.\ D {\bf 84}, 083519 (2011)
	[arXiv:1106.6025 [hep-th]].
	
\bibitem{NoScaleInflation}
W.~Buchmuller, C.~Wieck and M.~W.~Winkler,
Phys.\ Lett.\ B {\bf 736} (2014) 237
doi:10.1016/j.physletb.2014.07.024
[arXiv:1404.2275 [hep-th]].
T.~Li, Z.~Li and D.~V.~Nanopoulos,
JCAP {\bf 1402} (2014) 028
doi:10.1088/1475-7516/2014/02/028
[arXiv:1311.6770 [hep-ph]].
S.~Antusch, M.~Bastero-Gil, K.~Dutta, S.~F.~King and P.~M.~Kostka,
JCAP {\bf 0901} (2009) 040
doi:10.1088/1475-7516/2009/01/040
[arXiv:0808.2425 [hep-ph]].
P.~Binetruy and M.~K.~Gaillard,
Phys.\ Rev.\ D {\bf 34} (1986) 3069.
doi:10.1103/PhysRevD.34.3069
K.~Enqvist, D.~V.~Nanopoulos and M.~Quiros,
Phys.\ Lett.\  {\bf 159B} (1985) 249.
doi:10.1016/0370-2693(85)90244-8.
A.~Addazi, S.~V.~Ketov and M.~Y.~Khlopov,
Eur.\ Phys.\ J.\ C {\bf 78} (2018) no.8,  642
doi:10.1140/epjc/s10052-018-6111-7
[arXiv:1708.05393 [hep-ph]].

	\bibitem{Ellis:2013xoa}
	J.~Ellis, D.~V.~Nanopoulos and K.~A.~Olive,
	Phys.\ Rev.\ Lett.\  {\bf 111} (2013) 111301
	[arXiv:1305.1247 [hep-th]].


	\bibitem{Ellis:2013nxa}
	J.~Ellis, D.~V.~Nanopoulos and K.~A.~Olive,
	JCAP {\bf 1310} (2013) 009
	doi:10.1088/1475-7516/2013/10/009
	[arXiv:1307.3537 [hep-th]].
	
		
	\bibitem{Ellis:2013nka}
	J.~Ellis, D.~V.~Nanopoulos and K.~A.~Olive,
	Phys.\ Rev.\ D {\bf 89} (2014) no.4,  043502
	doi:10.1103/PhysRevD.89.043502
	[arXiv:1310.4770 [hep-ph]].

	
		
	

\bibitem{Romao:2017uwa}
  M.~C.~Romao and S.~F.~King,
  JHEP {\bf 1707} (2017) 033
  doi:10.1007/JHEP07(2017)033
  [arXiv:1703.08333 [hep-ph]].

	\bibitem{Wess:1974tw}
	J.~Wess and B.~Zumino,
	Nucl.\ Phys.\ B {\bf 70} (1974) 39.
	doi:10.1016/0550-3213(74)90355-1.
	
	\bibitem{Croon:2013ana}
	D.~Croon, J.~Ellis and N.~E.~Mavromatos,
	Phys.\ Lett.\ B {\bf 724} (2013) 165
	doi:10.1016/j.physletb.2013.06.016
	[arXiv:1303.6253 [astro-ph.CO]].
	
	\bibitem{Ellis:1984bm}
	J.~R.~Ellis, C.~Kounnas and D.~V.~Nanopoulos,
	Nucl.\ Phys.\ B {\bf 247} (1984) 373.
	doi:10.1016/0550-3213(84)90555-8
	
	



	
			
			
	
	\bibitem{Kallosh:2013xya}
	R.~Kallosh and A.~Linde,
	JCAP {\bf 1306} (2013) 028
	doi:10.1088/1475-7516/2013/06/028
	[arXiv:1306.3214 [hep-th]].
	
		\bibitem{No-scale}
		J.R. Ellis, K. Enqvist, D.V. Nanopoulos, K.A. Olive and M. Srednicki, SU(N; 1) in
		ation,
		Phys. Lett. B 152 (1985) 175 [Erratum ibid. B 156 (1985) 452].A.~S.~Goncharov and A.~D.~Linde,
		``A Simple Realization Of The Inflationary Universe Scenario In Su(1,1) Supergravity,''
		Class.\ Quant.\ Grav.\  {\bf 1} (1984) L75.
		doi:10.1088/0264-9381/1/6/004
		
	\bibitem{Ellis:2010jb}
	J.~Ellis, A.~Mustafayev and K.~A.~Olive,
	Eur.\ Phys.\ J.\ C {\bf 69} (2010) 219
	doi:10.1140/epjc/s10052-010-1400-9
	[arXiv:1004.5399 [hep-ph]].
	
	\bibitem{Ellis:2015xna}
	J.~Ellis, M.~A.~G.~Garcia, D.~V.~Nanopoulos and K.~A.~Olive,
	Class.\ Quant.\ Grav.\  {\bf 33} (2016) no.9,  094001
	doi:10.1088/0264-9381/33/9/094001
	[arXiv:1507.02308 [hep-ph]].
	J.~Ellis,
	doi:10.1142/9789813226609\_0004
	
	
	\bibitem{Ibe:2007}
	M. Ibe, T. Moroi and T. T. Yanagida, Phys. Lett. B 644, 355 (2007) [hepph/
0610277]; M. Ibe and T. T. Yanagida, Phys. Lett. B 709, 374 (2012)
[arXiv:1112.2462 [hep-ph]]; M. Ibe, S. Matsumoto and T. T. Yanagida, Phys. Rev.
D 85, 095011 (2012) [arXiv:1202.2253 [hep-ph]].




	\bibitem{Dudas:2012wi}
	E.~Dudas, A.~Linde, Y.~Mambrini, A.~Mustafayev and K.~A.~Olive,
	Eur.\ Phys.\ J.\ C {\bf 73} (2013) no.1,  2268
	doi:10.1140/epjc/s10052-012-2268-7
	[arXiv:1209.0499 [hep-ph]].
	
	\bibitem{Evans:2013lpa}
	J.~L.~Evans, M.~Ibe, K.~A.~Olive and T.~T.~Yanagida,
	Eur.\ Phys.\ J.\ C {\bf 73} (2013) 2468
	doi:10.1140/epjc/s10052-013-2468-9
	[arXiv:1302.5346 [hep-ph]].
	
	\bibitem{Evans:2013dza}
	J.~L.~Evans, K.~A.~Olive, M.~Ibe and T.~T.~Yanagida,
	Eur.\ Phys.\ J.\ C {\bf 73} (2013) no.10,  2611
	doi:10.1140/epjc/s10052-013-2611-7
	[arXiv:1305.7461 [hep-ph]].
	
	\bibitem{Evans:2014xpa}
	J.~L.~Evans and K.~A.~Olive,
	Phys.\ Rev.\ D {\bf 90} (2014) no.11,  115020
	doi:10.1103/PhysRevD.90.115020
	[arXiv:1408.5102 [hep-ph]].
	
	\bibitem{Evans:2014pxa}
	J.~L.~Evans, M.~Ibe, K.~A.~Olive and T.~T.~Yanagida,
	Phys.\ Rev.\ D {\bf 91} (2015) 055008
	doi:10.1103/PhysRevD.91.055008
	[arXiv:1412.3403 [hep-ph]].
	
	\bibitem{Evans:2015bxa}
	J.~L.~Evans, N.~Nagata and K.~A.~Olive,
	Phys.\ Rev.\ D {\bf 91} (2015) 055027
	doi:10.1103/PhysRevD.91.055027
	[arXiv:1502.00034 [hep-ph]].
	
\bibitem{Dine:1992}
M. Dine and D. MacIntire, Phys. Rev. D 46, 2594 (1992) [hep-ph/9205227]; L. Randall
and R. Sundrum, Nucl. Phys. B 557, 79 (1999) [arXiv:hep-th/9810155]; G. F. Giudice, M. A. Luty, H. Murayama and R. Rattazzi, JHEP 9812, 027 (1998) [arXiv:hepph/9810442]; J. A. Bagger, T. Moroi and E. Poppitz, JHEP 0004, 009 (2000)
[arXiv:hep-th/9911029]; P. Binetruy, M. K. Gaillard and B. D. Nelson, Nucl. Phys.
B 604, 32 (2001) [arXiv:hep-ph/0011081].

	\bibitem{Giudice:1988yz}
	G.~F.~Giudice and A.~Masiero,
	Phys.\ Lett.\ B {\bf 206} (1988) 480.
	doi:10.1016/0370-2693(88)91613-9
	E.~Dudas, Y.~Mambrini, A.~Mustafayev and K.~A.~Olive,
	Eur.\ Phys.\ J.\ C {\bf 72} (2012) 2138
	Erratum: [Eur.\ Phys.\ J.\ C {\bf 73} (2013) 2430]
	doi:10.1140/epjc/s10052-012-2138-3, 10.1140/epjc/s10052-013-2430-x
	[arXiv:1205.5988 [hep-ph]].

\bibitem{Evans:2019oyw}
J.~L.~Evans, N.~Nagata and K.~A.~Olive,
arXiv:1902.09084 [hep-ph].
\end{thebibliography}
\end{document}